\documentclass{article}
\usepackage{amsmath,amsthm}
\usepackage{graphicx}
\usepackage{hyperref}

\makeatletter
\newtheorem*{rep@theorem}{\rep@title}
\newcommand{\newreptheorem}[2]{%
	\newenvironment{rep#1}[1]{%
		\def\rep@title{#2 \ref{##1}}%
		\begin{rep@theorem}}%
		{\end{rep@theorem}}}
\makeatother

\newenvironment{lemma-repeat}[1]{\begin{trivlist}
		\item[\hspace{\labelsep}{\bf\noindent Lemma \ref{#1} }]\em }%
	{\end{trivlist}}
\newenvironment{theorem-repeat}[1]{\begin{trivlist}
		\item[\hspace{\labelsep}{\bf\noindent Theorem \ref{#1} }]\em }%
	{\end{trivlist}}

\newcommand{\qedsymb}{\qed}
\newenvironment{proofof}[1]{\begin{trivlist}
		\item[\hspace{\labelsep}{\bf\noindent Proof of #1: }]
	}{\qedsymb\end{trivlist}}

\newtheorem{theorem}{Theorem}[section]

\newtheorem{lemma}[theorem]{Lemma}

\newtheorem{definition}[theorem]{Definition}
\newtheorem{remark}[theorem]{Remark}
\newcounter{challenge}[section]
\newenvironment{challenge}[1][]{\refstepcounter{challenge}\par\medskip
	Challenge~\thechallenge. #1 \rmfamily}{\medskip}

\newcommand{\remove}[1]{}

\usepackage{amsfonts}
\usepackage[linesnumbered,vlined]{algorithm2e}
\graphicspath{ {./images/} }

\theoremstyle{remark}

\theoremstyle{definition}

\newcommand{\ip}[1]{\left}

\newcommand{\Alist}{\texttt{LIST}\xspace}
\newcommand{\Aarb}{\texttt{ARB-LIST}\xspace}

\newcommand{\congest}{\ensuremath{\mathsf{CONGEST}}\xspace}
\newcommand{\qcongest}{\ensuremath{\mathsf{QUANTUM~CONGEST}}\xspace}
\newcommand{\clique}{\ensuremath{\mathsf{CONGESTED~CLIQUE}}\xspace}

\newcommand{\mainTheoremText}{For all $p \geq 4$, there exists an algorithm for $K_p$-listing in the \congest model which completes in $\tilde{O}(n^{3/4} + n^{p/(p+2)})$ rounds, w.h.p..}

\newcommand{\specialCaseThoerem}{There exists an algorithm for $K_4$-listing in the \congest model which completes in $\tilde{O}(n^{2/3})$ rounds, w.h.p..}

\newcommand{\cliqueTheorem}{For all $p \geq 3$, there exists an algorithm for $K_p$-listing in the \clique model which completes in $\tilde{\Theta}(1 + m/n^{1 + 2/p})$ rounds, w.h.p..}

\usepackage{fullpage}
\begin{document}
	\title{
		On Distributed Listing of Cliques
	}
	
	\date{}
	\author{
Keren Censor-Hillel\\
 Technion \\
\and
Fran\c{c}ois Le Gall\\
 Nagoya University \\
\and
Dean Leitersdorf\\
 Technion  
 }
%

\maketitle
	\begin{abstract}
		We show an $\tilde{O}(n^{p/(p+2)})$-round algorithm in the \congest model for \emph{listing} of $K_p$ (a clique with $p$ nodes), for all $p =4, p\geq 6$. For $p = 5$, we show an $\tilde{O}(n^{3/4})$-round algorithm. 
				
		For $p=4$ and $p=5$, our results improve upon the previous state-of-the-art of $O(n^{5/6+o(1)})$ and $O(n^{21/22+o(1)})$, respectively, by Eden et al. [DISC 2019]. For all $p\geq 6$, ours is the first sub-linear round algorithm for $K_p$ listing.
		
		We leverage the recent expander decomposition algorithm of Chang et al. [SODA 2019] to create clusters with a good mixing time. 
Three key novelties in our algorithm are: (1) we carefully iterate our listing process with coupled values of min-degree within the clusters and arboricity outside the clusters,  (2)  all the listing is done within the cluster, which necessitates new techniques for bringing into the cluster the information about \emph{all} edges that can potentially form $K_p$ instances with the cluster edges, and (3) within each cluster we use a sparsity-aware listing algorithm, which is faster than a general listing algorithm and which we can allow the cluster to use since we make sure to sparsify the graph as the iterations proceed. 

		As a byproduct of our algorithm, we show an \emph{optimal} sparsity-aware algorithm for $K_p$ listing, which runs in $\tilde{\Theta}(1 + m/n^{1 + 2/p})$ rounds in the \clique model. Previously, Pandurangan et al. [SPAA 2018], Chang et al. [SODA 2019], and Censor-Hillel et al. [TCS 2020] showed sparsity-aware algorithms for the case of $p = 3$, yet ours is the first such sparsity aware algorithm for $p \geq 4$.
		\end{abstract}
		
\section{Introduction}

The problem of listing cliques of size $p$, as well as many additional subgraph-related problems, is a fundamental problem that has been extensively studied in many computational settings. Given a subgraph $H$ and a graph $G$, the problem of $H$-listing (also referred to as \emph{enumeration}) requires that every node outputs a set of instances of $H$, such that the union of all outputs is the list of all instances of $H$ in $G$. 

We achieve $K_p$ listing in the \congest model\footnote{In the \congest model, the $n$-node graph $G$ is the communication graph and messages of $O(\log{n})$ bits can be sent in synchronous rounds.} in a sub-linear number of $\tilde{O}(n^{p/(p+2)})$ rounds, for all $p = 4, p \geq 6$, and in $\tilde{O}(n^{3/4})$ rounds for $K_5$.\footnote{We use the $\tilde{O}(\cdot)$ notation to hide polylogarithmic multiplicative factors. All the logarithms in the paper are in base 2.} 

The first breakthrough in this area was the sub-linear algorithm for $K_3$ listing of Izumi and Le Gall \cite{Izumi+PODC17}, which was followed by the insightful algorithms of Chang et al.~\cite{Chang+SODA19} and Chang and Saranurak~\cite{Chang+PODC19} who brought the complexity down to a tight $\tilde{O}(n^{1/3})$ number of rounds. When $p \geq 4$, many additional challenges arise for $K_p$ listing, with some obstacles already appearing at $p=4$, and others at $p\geq 5$. Recently, Eden et al.~\cite{Eden+DISC19} presented the first sub-linear algorithms for $K_4, K_5$ listing, running in $O(n^{5/6 + o(1)})$ and $O(n^{21/22 + o(1)})$ rounds, respectively, overcoming some significant obstacles. 

For $p \geq 6$, no sub-linear time algorithms were known for $K_p$ listing prior to our work. 

Our algorithm relies on a new set of techniques which simultaneously solve $K_p$ listing in a sub-linear number of rounds, for all $p \geq 4$. We leverage the recent expander decomposition algorithm of Chang et al.~\cite{Chang+SODA19} to create clusters with a good mixing time. 
Three key novelties in our algorithm are: (1) we carefully iterate our listing process with coupled values of min-degree within the clusters and arboricity outside the clusters,  (2)  all the listing is done within the cluster, which necessitates new techniques for bringing into the cluster the information about \emph{all} edges that can potentially form $K_p$ instances with the cluster edges, and (3) within each cluster we use a sparsity-aware listing algorithm, which is faster than a general listing algorithm and which we can allow the cluster to use since we make sure to sparsify the graph as the iterations proceed.

The following is the formal statement of our main contribution.
\begin{theorem}
	\label{theorem:Kp}
	\mainTheoremText
\end{theorem}

Notice that for all $p \geq 6$, the $O(n^{p/(p+2)})$ term dominates. For the case of $K_4$, we are able to remove the first term and achieve an even faster algorithm which takes $\tilde{O}(n^{2/3})$, giving us the following.

\begin{theorem}
	\label{theorem:K4}
	\specialCaseThoerem
\end{theorem}

Nonetheless, for the lone case of $K_5$, the $O(n^{3/4})$ term remains and dominates the second. Most of the paper is devoted to proving Theorem~\ref{theorem:Kp}, and in Section~\ref{section:improvementsForK4} we show the modifications required in order to get rid of the first term for the case of $K_4$ and prove Theorem~\ref{theorem:K4}.

Notice that our results get closer to the lower bound  of $\tilde{\Omega}(n^{(p-2)/p})$ shown in Fischer et al.~\cite{Fischer+SPAA18}. \\

Lastly, we also present the following result in the \clique model.\footnote{In the \clique model, the $n$-node graph $G$ is the input graph and messages of $O(\log{n})$ bits can be sent in synchronous rounds between \emph{any} two nodes.}
\begin{theorem}
	\label{theorem:clique}
	\cliqueTheorem
\end{theorem}

Here, $m$ is the number of edges in the input graph. This algorithm is a byproduct our sparsity aware algorithm used in proving Theorem~\ref{theorem:Kp}, and so its formal proof is deferred to Section~\ref{section:clique}.

\subsection{The challenges}
The ingenious $K_3$ listing algorithms of~\cite{Chang+SODA19, Chang+PODC19}  construct and apply \emph{expander decompositions} which break up the input graph into dense clusters with good mixing times. Then, each cluster lists all the $K_3$ instances which have at least one edge within the cluster itself. When moving to $K_{p}$ listing with $p \geq 4$, a critical dissimilarity arises: a $K_{p \geq 4}$ instance with a single edge in a specific cluster can also have edges which are not incident to any of the cluster nodes, unlike in the $K_3$ case. 
This difference raises two main challenges which we address throughout the paper: 

\begin{challenge}
	\label{challenge:bringingInEdges}
	After applying the expander decomposition, for each cluster we need to ensure that any edge $e$ which participates in a $K_{p \geq 4}$ instance involving some edge inside the cluster, such that $e$ is not incident to any of the cluster nodes, is known to some node in the cluster.
\end{challenge}
\vspace{-3pt}

\begin{challenge}
	\label{challenge:listingInsideClusters}
	We need to perform the listing process efficiently within each cluster, despite the fact that after bringing edges into a cluster, the amount of information the cluster has to process can be substantially larger than the bandwidth available within the cluster.
\end{challenge}

In Eden et al.~\cite{Eden+DISC19}, the first challenge is tackled for the $K_4$ case. This is done by splitting the nodes outside a cluster into \emph{heavy} and \emph{light} nodes, where heavy nodes have the required bandwidth in order to send their entire neighborhood into the cluster, while light nodes do not have many neighbors \emph{inside the cluster} and thus can, with few queries to the cluster nodes, list all the $K_4$ which they share with the cluster nodes. This novel technique resolves the Challenge~\ref{challenge:bringingInEdges}. However, overcoming the second challenge is necessary for further improving the runtime.

In the cases of $K_{p\geq 5}$, both challenges remain, since unlike in $K_4$, there can be \emph{three nodes outside a cluster involved in a $K_{p\geq5}$ instance with a cluster edge}. Thus, now a \emph{light} node would also have to learn about edges outside the cluster, in order to determine if it is in a $K_{p\geq 5}$, incurring an overhead of too many rounds. For this reason, the algorithm for $K_5$ in~\cite{Eden+DISC19} takes a very different approach than the one they present for $K_4$. 

\subsection{Our approach}
The key ingredients of our approach for solving these challenges are \emph{controlling} the sparsity of the problem assigned to each cluster, and creating a \emph{sparsity-aware} algorithm based on a wide array of critical observations. Our result presents a unified algorithm which solves Challenge~\ref{challenge:bringingInEdges} in $\tilde{\Theta}(n^{3/4})$ rounds, regardless of the value of $p$, and then solves Challenge~\ref{challenge:listingInsideClusters} in $\tilde{\Theta}(n^{p/(p+2)})$ rounds. These guiding principles utilized in solving these challenges may turn useful for other subgraph related problems in the \congest model.

We first present how to overcome Challenge~\ref{challenge:listingInsideClusters}, since the solution for Challenge~\ref{challenge:bringingInEdges} relies on it.

~\\\textbf{Coping with Challenge~\ref{challenge:listingInsideClusters}: Controlling the \emph{bandwidth vs. problem size} ratio.} A necessary (though insufficient) requirement for speeding up the round complexity in the \congest model is ensuring that the bandwidth available to each cluster is proportional to the size of the problem assigned to it, that is, to the number of edges for which it must perform $K_p$ listing.

To see this, consider the case of $K_3$ in the \congest and the \clique models. The round complexity of $K_3$ listing in the \clique model is $\tilde{\Theta}(n^{1/3})$ rounds, as mentioned above. Nonetheless, as shown by Pandurangan et al.~\cite{Pandurangan+SPAA18} and by Censor-Hillel et al.~\cite{Censor+TCS20}, if the input graph is \emph{sparse}, it is possible to perform $K_3$ listing in $o(n^{1/3})$ rounds in the \clique model, and even in $O(1)$ rounds if $m = O(n^{5/3})$, where $m$ is the number of edges in $G$. Intuitively, for similar reasons, it should hold that using a \clique algorithm in a cluster with $k$ nodes in order to list all $K_3$ instances in an input graph with $\omega(k)$ nodes and $\omega(k^2)$ edges, should incur a round complexity which is $\omega(k^{1/3})$.

This intuition carries over to all $K_p$ and, as such, when using the expander decomposition, we should assign each cluster a $K_p$ listing problem where the number of input edges and the bandwidth available are closely related -- we ensure that the ratio between these values is at most $n/k$. 

Assigning a not-too-large listing problem to clusters was first done in \cite{Chang+SODA19} in order to get the $\tilde{O}(n^{1/2})$-round algorithm for $K_3$ listing in the \congest model, and we ensure this in the significantly more challenging case of $K_{p \geq 4}$. The reason this case is drastically more difficult is due to Challenge~\ref{challenge:listingInsideClusters} which applies only for $K_{p\geq 4}$ and not for $K_3$. 

It is therefore paramount to control the size of the problem given to each cluster. Each cluster is assigned a single task: \emph{to list all the $K_p$ which contain at least one edge inside the cluster}. Each such $K_p$ can have three types of edges: edges inside the cluster, edges crossing the cluster boundary (one node inside the cluster and one outside), and edges entirely outside the cluster, that touch two neighbors of the cluster. We achieve this control using the following strategies.

~\\\textbf{Coping with Challenge~\ref{challenge:listingInsideClusters}: Keeping minimal degree and arboricity close together.}
Our key approach in order to ensure that the number of edges of the first, second, and third types is proportional to the bandwidth used inside the clusters, is to make sure that the \emph{minimal degree} inside the clusters is \emph{always very close} to the \emph{arboricity} of the entire graph. 

We do this by employing two, nested, iterative processes. The outer process decreases the arboricity and the inner processes decreases the average degree in the graph. These new iterative procedures are the key concepts of our algorithm which control the ratio between the computation bandwidth and the problem size.

We get two major advantages by having these iterative processes. First, we promise that the ratio between the number of edges brought into the cluster and the number of edges inside the cluster is roughly $n/k$, as required. Second, we guarantee that the number of edges inside the cluster is very close to the bandwidth that we actually use for routing, 

which is the product of the number of nodes in the cluster and the minimal degree within the cluster. This allows us to avoid the partitioning of vertices into degree classes that is done in~\cite{Chang+SODA19, Chang+PODC19}. 

~\\\textbf{Coping with Challenge~\ref{challenge:listingInsideClusters}: Sparsity-aware listing.} As stated, controlling the ratio between bandwidth and problem size is a necessary condition for fast $K_p$ listing, yet, this condition is insufficient on its own. Therefore, we leverage our approach of decreasing arboricity to argue that the graph becomes sparse as the algorithm progresses, which enables us to utilize an efficient sparsity aware algorithm. To this extent, we create a novel \clique-style sparsity-aware $K_p$ listing algorithm for all $p \geq 3$. Notice that previously \cite{Pandurangan+SPAA18, Censor+TCS20, Chang+SODA19} showed algorithms with similar properties, yet only for $p=3$. Further, in Section~\ref{section:clique}, we prove that this algorithm can also be used in the \clique model itself as a general sparsity aware algorithm.

~\\\textbf{Coping with Challenge~\ref{challenge:bringingInEdges}: Delaying treatment of bad edges to future iterations.}  
Finally, we need to ensure that all the edges outside the cluster which could possibly generate a $K_{p\geq 4}$  instance with some edge in the cluster become known in the cluster. This property has not been previously achieved, and is the key for what allows our algorithm to work for $K_p$ of all $p \geq 4$, simultaneously. To this extent, we enhance the technique of considering \emph{heavy} and \emph{light} nodes as first defined by Eden et al.~\cite{Eden+DISC19}. Nodes outside the cluster are classified as either heavy or light, depending on how many neighbors they have within the cluster.

In~\cite{Eden+DISC19}, heavy nodes send their neighbors into the cluster, while light nodes list $K_4$ instances themselves.

Our algorithm brings all neighboring edges into the cluster itself. The huge challenge with light nodes is that they may have much information to send into the cluster, but only a small bandwidth into the cluster to use for sending this information. 

Here, we observe that since light nodes have few cluster neighbors, then, on average, most of the cluster nodes should have few light neighbors outside the cluster. Thus, we detect problematic nodes within the clusters (those which have too many light neighbors) and move the edges inside the cluster which are connected to them to the next iterations of the algorithm. This ensures that each remaining cluster node has few enough light neighbors, ensuring that the cluster does not need to learn many edges involving light nodes and thus all those edges can be sent efficiently into the cluster.

We mention that the triangle listing algorithm of~\cite{Chang+SODA19} also delays treatment of some edges to future iterations. However, these are different edges and this is done for different reasons than ours. In the triangle listing algorithm, the edges are moved in order to bound the number of edges crossing the cluster boundary that need to be \emph{processed} because they are a part of the input for the cluster (but they are already known to the cluster). In our algorithm, the reason for moving edges is in order to bound the number of light neighbors that a cluster node has, so that we bound the amount of information it has has to \emph{learn}.

Lastly, we must also ensure that after sending the information from outside the cluster into it, no single node in the cluster becomes responsible for too many edges from outside the cluster, since otherwise it would not be possible to perform the \emph{sparsity-aware} algorithm efficiently. Therefore, we leverage the guarantees we maintain regarding the arboricity of the graph during our iterations in order to be able to generate a \emph{load-balanced} partition of the edges from outside the cluster. 

\subsection{Related Work}
As mentioned, the first sublinear algorithm for clique listing in the \congest model is due to Izumi and Le Gall~\cite{Izumi+PODC17}, who showed a $\tilde{O}(n^{3/4})$-round algorithm for listing triangles. This was followed by a $\tilde{O}(n^{1/2})$-round algorithm of Chang et al.~\cite{Chang+SODA19}  and a $\tilde{O}(n^{1/3})$-round algorithm of Chang and Saranurak~\cite{Chang+PODC19}. The latter is tight up to polylogarithmic factors, due to a matching lower bound by Pandurangan et al.~\cite{Pandurangan+SPAA18} and Izumi and Le Gall~\cite{Izumi+PODC17}. This is also the current state-of-the-art for triangle \emph{detection}, requiring that some node indicates the existence of a triangle if there is such, for which it is only known that a single round does not suffice, by either deterministic or randomized algorithms, due to Abboud et al.~\cite{Abboud+arxiv17} and Fischer et al.~\cite{Fischer+SPAA18}, respectively. 

Recently, a result by Huang et al.~\cite{Huang+SODA20} showed that it is possible to solve triangle listing in $O(\Delta/\log n + \log \log n)$ rounds in the \congest model, where $\Delta$ denotes the maximal degree in the graph. This is the first algorithm which is sub-linear in $\Delta$ for this problem. In fact, their solution also holds for the \emph{more difficult} version of triangle listing, known as \emph{local} triangle listing, where each triangle needs to be reported by at least of one of its three member nodes. This problem is known to take $\Omega(\Delta / \log n)$ rounds due to \cite{Izumi+PODC17}.

For cliques of size $p\geq 4$, the first sublinear algorithms were given by Eden et al.~\cite{Eden+DISC19}, who showed that $K_4$ can be listed in $O(n^{5/6+o(1)})$ rounds and that $K_5$ can be listed in $O(n^{21/22+o(1)})$ rounds. 

Fischer et al.~\cite{Fischer+SPAA18} show a lower bound of $\tilde{\Omega}(n^{(p-2)/p})$ for $K_p$ listing. For the detection version of cliques the only lower bound known is due to Czumaj and Konrad~\cite{Czumaj+DISC18}, who show that $\tilde{\Omega}(n^{1/2})$ rounds are needed for $K_p$ detection for all $4 \leq p \leq n^{1/2}$ and that $\tilde{\Omega}(n/p)$ rounds are needed for $K_p$ detection for all $p \geq n^{1/2}$.

The core method of using an \emph{expander decomposition} has been widely used before, but was first given for the \congest model by Chang et al.~\cite{Chang+SODA19}. A different decomposition was given in~\cite{Chang+PODC19}, both for listing triangles. Eden et al.~\cite{Eden+DISC19} use this decomposition to create another type of \emph{layered decomposition}, which they use for $K_4$ and $K_5$ listing, as well as for showing how to list arbitrary $p$-node subgraphs in $O(n^{2-2/(3p+1)+o(1)})$ rounds, for constant $p$.

For cycles, Drucker et al.~\cite{Drucker+PODC14} showed that for fixed $p \geq 4$, $C_p$ detection requires $\Omega(ex(n,C_p)/n))$ rounds, where $ex(n,H)$ is the Turan number that counts the maximum number of edges that an $n$-node graph can have without containing an isomorphic subgraph to $H$. For odd values of $p$ this implies a lower bound of $\tilde{\Omega}(n)$, while for $p=4$ it implies a lower bound of $\tilde{\Omega}(n^{1/2})$. The latter was then extended by Korhonen and Rybicki~\cite{Korhonen+OPODIS17} who make the $\tilde{\Omega}(n^{1/2})$ lower bound apply for any even value of $p$. They also show an algorithm for $C_p$ that completes within a linear number of rounds for any constant $p$, implying that for constant odd values the complexity is $\tilde{\Theta}(n)$. For even values, Fischer et al.~\cite{Fischer+SPAA18} showed that $C_{2p}$ can be solved in $O(n^{1-1/(p(p-1))})$ rounds, which was later improved by Eden et al.~\cite{Eden+DISC19} to $\tilde{O}_p(n^{1-2/(p^2-p+2)})$ rounds for odd $p \geq 3$, and at most $\tilde{O}_p(n^{1-2/(p^2-2p+4)})$ rounds for even $p \geq 4$.\footnote{The $O_p(\cdot)$ notation refers to the $O(\cdot)$ notation, while treating $p$ as a constant in terms of multiplicative factors to the round complexity.}

Even et al.~\cite{Even+DISC17} and~\cite{Korhonen+OPODIS17}  also show algorithms for detection of trees and additional subgraphs.
Additional lower bounds for subgraph detection are given in~\cite{Fischer+SPAA18}, showing a lower bound of $\Omega(n^{2-1/p} /p)$ rounds for a family of graphs $H_p$ with $p$ nodes. Additional lower bounds are given by Gonen and Oshman in~\cite{Gonen+OPODIS17}.

\section{Sub-linear $K_p$-listing, for $p\geq 4$}
\label{section:listing}
\subsection{Preliminaries}
Throughout the algorithm, we use the expander decomposition of \cite{Chang+SODA19},\footnote{We note that our algorithmic techniques are fundamentally \emph{incompatible} with the improved expander decomposition seen in \cite{Chang+PODC19}, due to the fact that we heavily rely on a result related to the arboricity of parts of the decomposition -- a notion which is central to \cite{Chang+SODA19} but which exhibits an obstacle towards triangle listing and hence is successfully removed in \cite{Chang+PODC19}.} and therefore we define here notation which relates to this. We begin by defining the notion of clusters, which are components that have a lower bound on the degrees of their vertices as well as a small \emph{mixing time}, where \emph{mixing time} roughly denotes the number of rounds required for a random walk to reach the stationary distribution. 

\begin{definition}[Clusters~\cite{Chang+SODA19}]
	\label{definition:cluster}
	Given a graph $G=(V, E)$, a set $V' \subseteq V$ is an \emph{$n^\delta$-cluster w.r.t $E' \subseteq E$}, if it is a maximal connected component in the graph $G'=(V, E')$ and it has the following properties: (1) each node $v' \in V'$ has $deg_{E'}(v) = \Omega(n^\delta)$, and (2) the mixing time of $V'$ in $G'$ is $O(polylog(n))$.
\end{definition}

Our algorithm relies on having a decomposition of the graph into such clusters, defined as follows.

\begin{definition}[$\delta$-Expander Decomposition~\cite{Chang+SODA19}]
	\label{definition:decomposition}
	Given a graph $G=(V, E)$ and $0<\delta< 1$, a \emph{$\delta$-decomposition of $G$} is a partition of its edge set into $E = E_m \cup E_s \cup E_r$, such that the following hold:
	\begin{itemize}
		\item $E_m$ is such that each maximal connected component w.r.t to $E_m$ that includes more than one node is an $n^\delta$-cluster. Further, for each cluster in $E_m$, there is a unique identifier known to all nodes of the cluster, and each node knows which of its edges are in $E_m$ and to which cluster it belongs. 
		
		\item The arboricity of the subgraph induced by $E_s$ is at most $n^\delta$. Further, there exists an orientation of the edges such that $E_s = \cup_{v\in V} E_{s, v}$, where $E_{s, v}$ is the set of edges of $E_s$ oriented \emph{away} from $v$, and $|E_{s, v}| \leq n^\delta$. Each node $v$ knows which of its edges are in $E_{s,v}$.
		
		\item $|E_r| \leq |E| / 6$.
	\end{itemize}
	
\end{definition}

A  $\delta$-expander decomposition has been constructed by Chang et al.~\cite{Chang+SODA19}, giving the following.

\begin{theorem}[$\delta$-Decomposition Construction~\cite{Chang+SODA19}]
	\label{theorem:decomposition}
	There exists an algorithm for constructing a $\delta$-expander decomposition in the \congest model which completes in $\tilde{O}(n^{1- \delta})$ rounds. 
\end{theorem}

The algorithm given in~\cite{Chang+SODA19} also promises that each cluster has an ID that is known to all cluster nodes.

Our algorithms rely on the ability to perform quick routing within the clusters in the expander decomposition. We use the following theorem which follows from the routing algorithms of~\cite{Ghaffari+PODC17} and~\cite{Ghaffari+DISC18}. This theorem appears as Theorem 4.1 in \cite{Chang+SODA19} and is discussed more in-depth in Section 3 of \cite{Chang+PODC19}.

\begin{theorem}{Intra-Component Routing.}
	\label{theorem:generalRouting}
	Let $G = (V, E)$ be a graph and $0 < \delta < 1$. Let $C$ be an $n^\delta$-cluster in $G$. If every node in $C$ has at most $O(n^\delta\cdot2^{O(\sqrt{\log n})})$ messages it needs to send and receive, then there exists an algorithm in the \congest model that routes all messages within $C$ in $\tilde{O}(2^{O(\sqrt{\log n})})$ rounds.\footnote{The constant  factors used in the exponents are different (personal communication with the authors of \cite{Chang+PODC19}). That is, the statement holds if each node wants to send and receive $O(n^\delta\cdot2^{c_1\sqrt{\log n}})$ messages in a total of $\tilde{O}(2^{c_2\sqrt{\log n}})$ rounds, for some constants $c_1, c_2$. Thus, direct usage of this theorem would negatively impact our final results and would add a factor of $n^{o(1)}$ to the round complexities of the $K_p$ listing algorithms we show. However, similarly to the discussion found in Section 3 of \cite{Chang+PODC19}, in our case it is also possible to overcome this extra term due to a trade-off present in the routing algorithm, since our final round complexities are $\Omega(n^{1/3})$.} 
\end{theorem}

We emphasize that Theorem~\ref{theorem:generalRouting} only uses the edges of $C$ for routing, thus one can route in multiple clusters in parallel.
Further, Lemma 4.1 in \cite{Chang+SODA19}, also provides us with the following Lemma~\ref{lemma:IDs} which is used in the final part of our algorithm.

\begin{lemma}{Intra-Component ID Assignment.}
	\label{lemma:IDs}
	Let $G = (V, E)$ be a graph and $0 < \delta < 1$, and $C_1, \dots, C_q$ the $n^\delta$-clusters in the above expander decomposition of $G$ w.r.t. $\delta$. Then it is possible in $O(polylog(n))$ rounds, in the \congest model, to compute new ID assignments, $C \rightarrow \{1, \dots, |C| \}$, for each $C$ out of $C_1, \dots, C_q$, in parallel.
\end{lemma}

We note the following remark which splits $K_p$ listing into two cases, when $p = \omega(\log n)$ and when $p = O(\log n)$.

\begin{remark}
	\label{remark:pAtMostLogn}
	Notice that for $p = \omega(\log n)$, the lower bound for $K_p$ listing is $\tilde{\Omega}(n^{(p-2)/p}) = \tilde{\Omega}(n^{1 - 2/p})  = \tilde{\Omega}(n)$, and, therefore, for these values of $p$, one can trivially list all $K_p$ in $\tilde{\Theta}(n)$ rounds by having each node broadcast its neighborhood. Thus, we can assume for the rest of our algorithm that $p = O(\log n)$.
\end{remark}

Lastly, we require the following \emph{input partitioning} lemma, which appears as Lemma 4.2 in \cite{Chang+SODA19}. 
\begin{lemma}{\cite[Lemma 4.2]{Chang+SODA19}}
	\label{lemma:partition}

Given a graph with $\bar{m}$ edges and $\bar{n}$ vertices, generate a subset $S$ by letting each node join $S$ independently with probability $q$. Suppose that the maximum degree is $\Delta \leq \bar{m}q/20\log\bar{n}$ and $q^2\bar{m} \geq 400 \log^2{\bar{n}}$. Then, with probability at least $1 - 10(\log\bar{n})/\bar{n}^5$, the number of edges in the subgraph induced by $S$ is at most $6q^2\bar{m}$.
\end{lemma}

We are now ready to prove our main contribution.

\begin{theorem-repeat}{theorem:Kp}
	\mainTheoremText
\end{theorem-repeat}

\subsection{Iteratively decreasing the arboricity}
One of the main ingredients in proving Theorem~\ref{theorem:Kp} is an algorithm which removes edges from the graph in order to decrease its arboricity, while listing $K_p$ instances that contain at least one of the removed edges. This is formally given as follows.
\begin{theorem}
	\label{theorem:Kp_helper}
	For all $p \geq 4$, there exists an algorithm denoted \Alist, which, given a graph $G = (V, E)$ with arboricity at most $A$, along with an orientation of its edges with a maximum out-degree of $A$, such that $n^{p/(p+2)} < A/(2 \log n)$, splits $E$ into two edge sets $E = \tilde{E}_m \cup \tilde{E}_s$, such that the arboricity in $\tilde{E}_s$ is at most $A / 2$, the edges of $\tilde{E}_s$ are oriented with a maximum out-degree of at most $A / 2$, and \Alist lists all $K_p$ instances in $G$ which have at least one edge in $\tilde{E}_m$. The algorithm completes in $\tilde{O}(n^{3/4} + n^{p/(p+2)})$ rounds.
\end{theorem}

For the following discussion, we assume that $A = n^d$, for some value of $d$, and denote by $\delta = d -  (1 + \log \log n) / \log n$. Notice that $n^\delta = A / (2\log n)$, and thus we can restate the theorem as having to ensure the arboricity of $\tilde{E}_s$ is at most $n^\delta \log n$. Our algorithm runs in  $O(n^{3/4 + d - \delta} + n^{p/(p+2) + d - \delta})$ rounds, which, due to the choice of $\delta$, is equivalent to $\tilde{O}(n^{3/4} + n^{p/(p+2)})$.

We use Theorem \ref{theorem:Kp_helper} iteratively on $\tilde{E}_s$ to prove Theorem~\ref{theorem:Kp}, as follows.
\begin{proofof}{Theorem~\ref{theorem:Kp}}
	
	The high-level approach of this proof is to use Theorem~\ref{theorem:Kp_helper} iteratively on a sequence of graphs with decreasing arboricity. Notice that all these graphs have \emph{the same node set}, and thus the value of $n$, the number of nodes in the graph, is well defined and does not change throughout the algorithm.
	
	We denote $G_0=G$, and let $\epsilon_0 =  (1 + \log \log n) / \log n$. We set $d_0 = 1$, which clearly gives that the arboricity in $G_0$ is at most $n^{d_0}$ and allows us to run Algorithm \Alist using $\delta_0 = 1 - \epsilon_0$. This creates a partition $\tilde{E}_{m,0},\tilde{E}_{s,0}$ and lists all $K_p$ instances which have at least one edge in $\tilde{E}_{m,0}$. This finishes within $\tilde{O}(n^{3/4 + d - \delta} + n^{p/(p+2) + d_0 -\delta_0}) = \tilde{O}(n^{3/4 + \epsilon_0} + n^{p/(p+2) + \epsilon_0})$ rounds.
	
	We are now left with the task of listing all $K_p$ instances in $G_0$ that have no edge in $\tilde{E}_{m,0}$. In other words, we need to list all $K_p$ instances which are fully contained in $\tilde{E}_{s,0}$. We define $G_1 = (V, \tilde{E}_{s,0})$ and notice that the arboricity in $G_1$ is at most $n^\delta \cdot \log n = n^{1 - \epsilon_0 + \log \log n / \log n}$. Therefore, we set $d_1 = 1 - \epsilon_0 + \log \log n / \log n$, $\epsilon_1 = 2\epsilon_0 - \log \log n / \log n$ and $\delta_1 = 1 - \epsilon_1 = 1-2\epsilon_0+ \log \log n / \log n$. We run Algorithm \Alist on $G_1$, which completes in $\tilde{O}(n^{3/4 + d - \delta} + n^{p/(p+2) + d_1 -\delta_1}) = \tilde{O}(n^{3/4 + \epsilon_0} + n^{p/(p+2) + \epsilon_0})$ rounds. Notice that this number of rounds is exactly the same as for the first invocation of Algorithm \Alist, since both $d_1$ and $\delta_1$ decrease by the same amount, $\epsilon_0 - \log \log n / \log n = 1 / \log n$.
	
	We continue iteratively applying Algorithm \Alist with $\epsilon_k = (k+1)\epsilon_0 - k \log \log n / \log n$, $\delta_k = 1-\epsilon_k$ and $d_k = \delta_{k} + \epsilon_0$. We do this for at most $k=1/(\epsilon_0 - \log \log n / \log n) = \log n$ iterations, as long as $\delta_k > p/(p+2)$ and $\delta_k > 3/4$. Once we get a $\delta_k \leq p/(p+2)$ or $\delta_k \leq 3/4$, we stop and observe that $d_k  = \delta_{k} + \epsilon_0$ and thus $d_k \leq p/(p+2) + \epsilon_0$ or $d_k \leq 3/4 + \epsilon_0$. At this stage, every node broadcasts its \emph{outgoing edges} to all its neighbors in $O(n^{d_k}) = O(n^{3/4 + \epsilon_0} + n^{p/(p+2) + \epsilon_0})$ rounds of communication, which ends the algorithm by listing all remaining $K_p$ instances (those that are contained in $G_k=(V,\tilde{E}_{s,k-1})$).
	
	To summarize the number of rounds, note that we iterate $k=O(\log n)$ times and in each iteration we run Algorithm \Alist in $\tilde{O}(n^{3/4 + \epsilon_0} + n^{p/(p+2) + \epsilon_0})$ rounds. Lastly, during the final step of the algorithm, the nodes broadcast whatever is left of their outgoing edges to their remaining neighbors, taking $O(n^{3/4 + \epsilon_0} + n^{p/(p+2) + \epsilon_0})$ rounds. Overall, since $\epsilon_0 =  (1 + \log \log n) / \log n$, the total number of rounds is $\tilde{O}(n^{3/4} + n^{p/(p+2)})$, completing the proof.
\end{proofof}

\subsection{Iterative arboricity-listing while decreasing the number of edges} 
We now show Algorithm \Alist from Theorem~\ref{theorem:Kp_helper}. We rely on the following procedure, which is the core of Algorithm \Alist.
\begin{theorem}
	\label{theorem:Kp_arb}
	For all $p \geq 4$, there exists an algorithm denoted \Aarb, which, given a graph with arboricity $n^d$ that is split to two edge sets, $E = E_s \cup E_r$, such that $E_s$ has arboricity $c\cdot n^\delta$, for a value $c$ and a value $\delta$ such that $p/(p+2) < \delta$, and $3/4 < \delta$, and $n^d = 2 \cdot n^\delta \cdot \log n$, along with an orientation of its edges with a maximum out-degree of $ c\cdot n^\delta$, splits the graph into three edge sets $\hat{E}_m, \hat{E}_s$ and $\hat{E}_r$, such that the arboricity in $\hat{E}_s$ is $(c + 1) \cdot n^\delta$, the edges of $\hat{E}_s$ are oriented with a maximum out-degree of $(c + 1) \cdot n^\delta$, the size of $\hat{E}_r$ is bounded by $|\hat{E}_r| \leq |E_r|/4$, and \Aarb lists all $K_p$ instances in $G(V, E)$ which have at least one edge in $\hat{E}_m$. The algorithm completes in $\tilde{O}(n^{3/4 + d - \delta} + n^{p/(p+2) + d - \delta})$ rounds.
\end{theorem}

Before proving Theorem~\ref{theorem:Kp_arb}, we show how it completes the proof of Theorem~\ref{theorem:Kp_helper}, as follows.
\begin{proofof}{Theorem~\ref{theorem:Kp_helper}}
	
	The high-level approach of this proof is to use Theorem~\ref{theorem:Kp_arb} iteratively on a sequence of graphs with a decreasing number of edges.
	
	We begin with the graph $G = (V, E)$, and denote $E_{s, 0} = \emptyset, E_{r, 0} = E$. We apply Algorithm \Aarb on this partition, and get a new partition $\hat{E}_{m,0},\hat{E}_{s,0}, \hat{E}_{r,0}$, such that the arboricity in $\hat{E}_{s,0}$ is $(0 + 1) \cdot n^\delta = n^{\delta}$, the edges of $\hat{E}_{s,0}$ are oriented with a maximum out-degree of $n^\delta$, the size of $\hat{E}_{r,0}$ is bounded by $|\hat{E}_{r,0}| \leq |E_{r, 0}|/4$, and \Aarb lists all $K_p$ instances which have at least one edge in $\hat{E}_{m,0}$. This finishes within $\tilde{O}(n^{3/4 + d - \delta} + n^{p/(p+2) + d - \delta})$ rounds.
	
	We are now left with the task of listing all $K_p$ instances in $G$ that have no edge in $\hat{E}_{m,0}$. In other words, we need to list all $K_p$ instances which are contained in $\hat{E}_{s,0}\cup \hat{E}_{r,0}$. We apply Algorithm \Aarb again with $E_{s, 1} = \hat{E}_{s,0}$ and $E_{r, 1} = \hat{E}_{r,0}$, getting the new $\hat{E}_{m,1},\hat{E}_{s,1}, \hat{E}_{r,1}$. Notice that \Aarb now lists all $K_p$ in $G(V, E_{s, 1} \cup E_{r, 1})$ which have at least one edge in $\hat{E}_{m,1}$. Thus, so far, \Aarb listed all $K_p$ in $G(V, E)$ with at least one edge in $\hat{E}_{m,1}$, since if any such $K_p$ has an edge in $E \setminus (E_{s, 1} \cup E_{r, 1}) = \hat{E}_{m, 0}$ then that $K_p$ would have already been listed by the first invocation of \Aarb. Thus, we can remove $\hat{E}_{m, 1}$ from the graph and continue with $\hat{E}_{s, 1}, \hat{E}_{r, 1}$. These two sets maintain that the arboricity of $\hat{E}_{s, 1} \leq 2 \cdot n^\delta$ (with a known corresponding orientation) and $|\hat{E}_{r, 1}| \leq |E_{r, 1}|/4 = |\hat{E}_{r, 0}|/4 \leq |E_{r, 0}|/16 = |E|/16$. 
	
	We continue iteratively applying Algorithm \Aarb on \\ $E_{s,k}, E_{r,k}$, obtaining that the arboricity of $\hat{E}_{s, k}$ is at most $(k + 1) \cdot n^\delta$ and that $|\hat{E}_{r, k}| \leq |E|/(4^{k+1})$. We do this for $k=\log n - 1$ iterations, until $|E_{r,k}| \leq |E|/(4^{\log{n}}) \leq (n \cdot ( n- 1)) / (4 ^ {\log n}) < 1$, which implies that $E_{r,k} = \emptyset$, and $E_{s,k}$ has an arboricity that is bounded by $n^{\delta} \cdot \log n$, as needed. During this iterative process, Algorithm \Aarb lists all the $K_p$ instances which have at least one edge in $E \setminus E_{s, k}$.
		
	To summarize the number of rounds, note that we iterate $k=O(\log{n})$ times and in each iteration we run Algorithm \Aarb in $\tilde{O}(n^{3/4 + d - \delta} + n^{p/(p+2) + d - \delta})$ rounds, giving the claimed complexity. 
\end{proofof}

\subsection{Algorithm \Aarb}
This subsection contains the proof of Theorem~\ref{theorem:Kp_arb}. 

The high-level idea of Algorithm \Aarb is running the expander decomposition with the given value $\delta$, on the graph $G = (V, E_r)$, producing $E_r = E'_m \cup E'_s \cup E'_r$. Then, we set $\hat{E}_s = E_s \cup E'_s$, select some $\hat{E}_m \subseteq E'_m$, and move the rest of the edges to $\hat{E}_{r} = E'_r \cup (E'_m \setminus \hat{E}_m)$. The choice of which edges to move is made so that it is easier to list all the instances of $K_p$ with at least one edge in $\hat{E}_m$ compared with listing all $K_p$ instances with at least one edge in $E'_m$. To make this precise, we say that an edge $e$ is a \emph{goal edge}, if the algorithm promises to list all instances of $K_p$ which contain $e$. Using this terminology, \Aarb sets $\hat{E}_m$ as goal edges, while edges that are moved from $E'_{m}$ to $\hat{E}_{r}$ are not goal edges (we call them \emph{bad edges}). 

However, if we simply remove edges from clusters in $E'_m$, we are no longer guaranteeing the properties of the cluster, such as an efficient mixing time. Thus, a crucial point for our algorithm to work is that we consider edges in $E_m\setminus\hat{E}_m$ as not being goal edges, but \emph{we still use them for communication in the clusters}.

We now show how to choose which edges to move and then how to list all the $K_p$ with at least one edge in $\hat{E}_m$. Both of these tasks are completed in $\tilde{O}(n^{3/4 + d - \delta} + n^{p/(p+2) + d - \delta})$ rounds. Notice that the initial expander decomposition takes $\tilde{O}(n^{1-\delta}) = \tilde{O}(n^{1/3})$, since $2/3 \leq p/(p+2) < \delta$. Thus, we achieve the required round complexity for Algorithm \Aarb.

\subsubsection{\textbf{Choosing bad edges and learning edges from outside the cluster}}
\label{subsec:bad-learn}

~\\Primarily, since we run the expander decomposition on $E_r$, we get that $|E'_r| \leq |E_r|/6$. Thus, in order to maintain the required guarantee that $|\hat{E}_r| \leq |E_r|/4$, we can move at most $(1/4 - 1/6) \cdot |E_r| = |E_r|/12$ edges from $E'_m$ to $\hat{E}_r$. This is thus the bound we strive to achieve on the number of edges moved. Nonetheless, since we do not focus on optimizing constant factors, we will show that the fraction of edges moved is $1/25 < 1/12$. 

Consider a single cluster $C$, and let $k$ be the number of nodes in $C$. Notice that $C$ has at least $k \cdot n^{\delta} / 2$ edges inside it due to the decomposition, yet at most $k \cdot n^d$ edges since the arboricity of the graph is $n^d$.

We now show how all edges that are not in $C$, and could potentially form $K_p$ instances with remaining goal edges in $C$, become known to nodes of $C$. These are edges between two nodes that are neighbors of the cluster. This process moves some edges from $E'_m$ to $\hat{E}_r$, in order to ensure that not too many edges from outside the cluster are brought into it.

~\\\textbf{Bad edges and learning edges from outside the cluster: }
At this stage, we wish to bound the amount of information which needs to enter the cluster by removing edges in $C$ which require too many edges from outside $C$ to be brought in. Every node $u \in C$ broadcasts to its neighbors outside $C$ a message that indicates that it is in cluster $C$ (recall that every node knows the ID of its cluster). Each neighbor $v$ of $C$ counts how many neighbors in $C$ it has, and denotes this value by $g_{v,C}$. If $g_{v,C} > n^{1/4}$, then $v$ is called a \emph{$C$-heavy node}, and otherwise it is called \emph{$C$-light}. 

Each $C$-heavy node $v$ has at most $n^d$ outgoing edges due to the arboricity of graph, and thus sends such edges into the cluster $C$, by sending each of its neighbors in $C$ a chunk of at most $O(n^{d - 1/4})$ of its outgoing edges. Note that this implies that each edge between two $C$-heavy nodes is thus known to some node $u \in C$.

For handling the edges of $C$-light nodes, we first need to account for nodes in $C$ which have too many $C$-light neighbors. For each node $u \in C$, we denote by $u_{light}$ the number of $C$-light neighbors it has. If $u_{light} > 100 \cdot n^{1/2} \cdot \log n$ then we say that $u$ is a \emph{bad node}. Every edge in $C$ that connects two bad nodes, is called a \emph{bad edge}, and is moved from $E'_m$ to $\hat{E}_r$ and thus is no longer a goal edge. We claim that there are at most a $|E'_m| / 25 \leq |E_r| / 25$  edges which are bad edges. To see why, note that the total number of edges between nodes in $C$ and $C$-light nodes is $n^{5/4}$, since there are at most $n$ $C$-light nodes, and each has at most $n^{1/4}$ neighbors in $C$. Therefore, there are at most $n^{5/4} / (100 \cdot n^{1/2} \cdot \log n) = n^{3/4}/(100 \log n) < k / (100 \log n)$ bad nodes, where the last inequality is since $k \geq n^\delta > n^{3/4}$. To now bound the number of edges removed, recall that the arboricity of the graph is $n^d$, and so there are at most $n^d \cdot k / (100 \log n)$ edges between bad nodes. On the other hand, the cluster has at least $n^\delta \cdot k / 2 = (n^d / (2 \log n)) k /2 = n^d \cdot k / (4 \log n)$ edges inside it, where the equality follows from the choice of $\delta$ w.r.t. $d$. Therefore, we removed at most $1/25$ of the cluster edges, and thus, summing across all clusters, we removed a total of $|E'_m| / 25$ edges, as claimed.

At this point, each \emph{good} node $u \in C$ has at most $\tilde{O}(n^{1/2})$ $C$-light neighbors. Each such node $u$ broadcasts its $C$-light neighbors to every neighbor $v$ that node $u$ has outside $C$, and receives from $v$ a list in which each item indicates whether a $C$-light neighbor $w$ of $u$ is also connected to $v$. Note that this implies that each edge between two neighbors of $C$ where one endpoint is $C$-light is thus known to some node $u \in C$. In Section~\ref{section:completenessProof}, we use this to show that $C$ knows all the graph edges which can potentially form a $K_p$ instance with at least one remaining goal edge in $C$.

We now bound the number of rounds we used so far, and the number of edges held by each node $u \in C$. Notice that each node $u\in C$ receives at most $O(n^{d - 1/4})$ edges from each neighbor $v \notin C$ of $u$. This is because if $v$ is $C$-heavy then it sends $u$ at most $O(n^{d - 1/4})$ edges when sending all its outgoing edges into the cluster, and, if $u$ is a good node, $v$ sends $u$ at most  $\tilde{O}(n^{1/2})$ additional edges when responding to $u$ after $u$ tells $v$ about all of its $C$-light neighbors (if $u$ is a bad node, no messages of the second type are sent). Thus, since $d \leq 1$, our runtime is bounded by $O(n^{3/4})$ for this step. Further, every node $u \in C$ receives at most $\tilde{O}(n^{d + 3/4})$ edges from outside the cluster.

\begin{remark}
	\label{remark:edges}
	We showed that each node $u \in C$ learns at most $\tilde{O}(n^{d + 3/4})$ edges that are completely outside the cluster. This is our desired bound since we know that $u$ can send and receive at least $\Omega(n^{\delta})$ messages quickly inside the cluster, and thus in $\tilde{O}(n^{d - \delta + 3/4})$ rounds, we later redistribute these edges inside the cluster in a \emph{load-balanced} way.
\end{remark}

\subsubsection{\textbf{Proving that all required edges are known to $C$}}
\label{section:completenessProof}

~\\ In this section we show that each edge outside of $C$ which can \emph{potentially} form a $K_p$ instance with at least one goal edge is known to some node in $C$. Let $H$ be some $K_p$ instance which contains at least one goal edge in $C$. Notice that all the other edges in $H$ can be either: inside $C$ (goal or non-goal edges), crossing the boundary of $C$, or entirely outside $C$. Each edge of the first two types is obviously known to some node in $C$, and thus it remains to show that all the edges outside $C$ in $H$ are known to some node or nodes in $C$. 

Notice that it suffices to show that any edge $e' = \{v, v'\}$ outside of $C$ which can form a $K_4$ with a goal edge $e = \{u, w\}$ of $C$ is known to some node in $C$, since if $e'$ is in a $K_p$ instance with $e$, then it is also in a $K_4$ instance with $e$.  Thus, let $H = \{u, w, v, v'\}$ be a $K_4$ instance such that $v, v' \notin C$ and $e = \{u, w\}$ is a goal edge of $C$. We show that $e' = \{v, v'\}$ is known to some node in $C$. 

\paragraph{Case 1: heavy-to-heavy edges} If both $v, v'$ are $C$-heavy, then the edge is directed away from one of them, and so that node sent $e'$ to one of its cluster neighbors.

\paragraph{Case 2: edge with a $C$-light endpoint} Assume w.l.o.g. that $v$ is $C$-light. Since $e$ is a goal edge of $C$, then at least one of its endpoints, w.l.o.g. assume it is $u$, is a good node. Thus, node $u$ sent the neighbor $v$ to $v'$ and $v'$ responded to $u$ that $e'$ exists and so node $u$  knows about $e'$.

\subsubsection{\textbf{Simulating a sparsity-aware \clique-style $K_p$-listing algorithm}}
\label{subsec:list}

What remains is to show our new sparsity-aware algorithm for $K_p$-listing, and prove that it can be executed efficiently within each cluster.
Let $C$ be a cluster with $k$ nodes denoted by $K = [k]$. Consider the set of edges that form an instance of $K_p$ with at least one goal edge in $C$. We have that each such edge is known to some node in $C$. We begin by running the algorithm from Lemma~\ref{lemma:IDs} for assigning new IDs in $[k]$ to the nodes of $C$, and from now on the nodes use these new IDs.

The main algorithmic ideas presented in this section are as follows. Prior to this step, every cluster $C$ reached a stage where the nodes of $C$ know all the information required in order to list all $K_p$ involving at least one edge in $C$. This was done by ensuring that each edge outside of $C$ which forms a $K_p$ involving at least one edge in $C$ is now known to at least one node in $C$. Now, the nodes of $C$ must efficiently communicate this information within the cluster in order to actually list all such $K_p$. Primarily, we reshuffle the edges known to the nodes of $C$ such that each node assumes responsibility for roughly the same amount of edges. Next, we create a randomized partition of the entire graph and show that the number of edges between any two parts of the partition are roughly the same. By doing so, we exploit the sparsity of the graph which we developed throughout the algorithm. Finally, each node in the cluster selects $p$ parts from the generated, randomized partition, and learns all the edges between these parts. By ensuring the every selection of $p$ parts is chosen by some node in the cluster, we guarantee that every $K_p$ with at least one edge inside $C$ is listed.

~\\\textbf{Reshuffling the edges:} In order to ensure a load-balanced and efficient execution of our sparsity aware algorithm later, we need all edges which are known to nodes in $C$ -- whether they are edges in $C$, crossing the cluster boundary of $C$, or completely outside $C$ -- to be grouped according to the node from which they are \emph{directed away from}. Concretely, for each node $v$ (whether $v \in C$ or $v \notin C$), we want to have a single node $u \in C$ which knows all of the edges directed away from $v$. Recall that since the graph has $n^d$ arboricity, and we know a corresponding orientation of the edges, then there are at most $n^d$ edges directed away from $v$. Therefore, each node $u \in C$ takes responsibility for $O(n/k)$ nodes in the graph. Precisely, the node with new ID $i \in [k]$ is responsible for the nodes whose (original) ID is in the range $[(i-1)\cdot n/k  + 1, i \cdot n/k]$. Using the routing algorithm of Theorem~\ref{theorem:generalRouting}, each node $u$ routes any edge which it originally receives from outside the cluster, \emph{and} any edge which is directed away from $u$ itself, to the node inside the cluster which are now responsible for the node from which that edge is outgoing. By Remark~\ref{remark:edges}, each node learns at most $\tilde{O}(n^{3/4 + d})$ edges from outside the cluster that must be routed. Further, since the arboricity of the graph is $n^d$, every node $u$ also has at most $n^d$ additional edges which are directed away from it and that must also be routed by $u$. At the end of the reshuffling, node $u$ is responsible for at most $O(n^d \cdot n/k) = O(n^{1/3 + d})$ edges (this is because $k \geq n^{\delta} > n^{p/(p+2)} \geq n^{2/3}$). Therefore, by Theorem~\ref{theorem:generalRouting}, the reshuffling procedure completes in $\tilde{O}(n^{3/4 + d - \delta})$ rounds.

~\\\textbf{Partitioning the graph:} We create a partition $\mathbb{V}$ of the entire graph, with $k^{1/p}$ roughly equally-sized parts. To do so, every node $u \in C$, for each node $w$ out of the $O(n/k)$ nodes outside the cluster which $u$ simulates, $u$ chooses uniformly at random which part in $\mathbb{V}$ the node $w$ joins. All in all, node $u$ makes $O(n/k)$ choices and broadcasts them to all nodes of $C$. This means that node $u$ sends and receives $O(k \cdot n/k) = O(n)$ messages, and thus this completes in $\tilde{O}(n^{1-\delta} + n^{o(1)}) = \tilde{O}(n^{1/3})$ rounds, using the algorithm from Theorem~\ref{theorem:generalRouting}, where we used $2/3 \leq p/(p+2) < \delta$.

Since there are at most $O(n^{1+d})$ edges in the graph, using a union bound with Lemma~\ref{lemma:partition} gives that, with high probability, the number of edges between any two parts in $\mathbb{V}$ is $O(n^{1 + d}/k^{2/p})$. Note that the conditions needed in Lemma~\ref{lemma:partition} are satisfied since $n^{1+d}/k^{1/p} \geq n^{3/4 + d} > n^{17/12} > n$, where the first inequality is since $k \leq n, 4 \leq p$ and the last inequality is since $d > p/(p+2) \geq 2/3$, and so obviously the maximal degree in the graph is below this value. 
 
~\\\textbf{Listing $K_p$ by learning graph edges:} Each node $u \in C$ is assigned, in a predetermined, balanced manner, $p$ parts in $\mathbb{V}$. The new IDs of the nodes are used to decide which parts they get, and since the nodes of $C$ have new IDs in $[k]$, each node can locally compute which parts were assigned to which node. Precisely, node $u$ views the $k^{1/p}$-radix representation of its new ID and uses the digits in the representation in order to determine the parts assigned to it.
Node $u$ then needs to learn all the edges between the parts that are assigned to it and list all instances of $K_p$ that it observes. Since the assignment is predetermined, any node $u$ in the cluster which holds an edge which node $w$ needs to learn, can send the edge to $w$. In order to do so in a load-balanced way, node $u$ sends such an edge to node $w$ only if in the orientation of the graph the edge is oriented away from one of the nodes which it simulates.

The number of messages each node receives is $O(p^2 n^{1+d}/k^{2/p})$. We know that $k > n^{p/(p+2)}$, and therefore, $O(p^2 n^{1+d}/k^{2/p}) = O(p^2 n^{1+d}/n^{(2/p)\cdot(p/(p+2))}) = O(p^2 n^{1+d - 2/(p+2)}) = O(p^2n^{p/(p+2) + d})$. It remains to show that each node also sends at most $O(p^2 n^{1+d}/k^{2/p}) = O(p^2n^{p/(p+2) + d})$ messages, and then by Theorem~\ref{theorem:generalRouting}, this part completes in $O(p^2n^{p/(p+2) + d - \delta})$ rounds. Notice that due to Remark~\ref{remark:pAtMostLogn}, we can hide the $O(p^2)$ term with the $\tilde{O}(\cdot)$ notation.

To show that node $u$ sends at most $O(p^2 n^{1+d}/k^{2/p})$ messages, recall that $u$ is responsible for at most $O(n^d \cdot n /k)$ edges in the graph. Each such edge needs to be sent to every node which selected the parts which contain both endpoints of that edge, and thus each edge is sent to at most $O(p^2k^{1-2/p})$ nodes\footnote{As stated above, the \emph{part assignment} is by the $k^{1/p}$-radix representation of the ID of a node. We denote by \emph{the $i^{th}$ part assigned to a node as the value of the $i^{th}$ digit of the $k^{1/p}$-radix representation of the ID of that node}. Let $A, B$ be two the parts in the partition which hold the endpoints of a given edge. There are $k^{1-2/p}$ nodes which were assigned $A, B$ as their first parts. This is because $k^{1-1/p}$ nodes are assigned their first part as $A$, and out of those nodes, a $k^{1/p}$ fraction are assigned $B$ as their second part. We then complete the bound by multiplying by $O(p^2)$ since we need to deliver to all nodes which are assigned $A, B$ and not just those assigned these parts as their first and second parts, respectively.}. Thus, $u$ sends at most $O(p^2 n^d \cdot n/k^{2/p})$ messages, as claimed.

\section{Faster $K_4$ Listing: in $\tilde{O}(n^{2/3})$ rounds}
\label{section:improvementsForK4}
We now present an additional improvement which overcomes the $O(n^{3/4})$ additive complexity in the previous algorithm for the case of $K_4$. We manage to completely overcome this challenge,  by not sending edges incident to $C$-light nodes into the cluster $C$, and thus we solve $K_4$ listing in $\tilde{O}(n^{p/(p+2)}) = \tilde{O}(n^{2/3})$ rounds. 

\begin{theorem-repeat}{theorem:K4}
	\specialCaseThoerem
\end{theorem-repeat}

\begin{proof}
In order to get the improved runtime for $K_4$ listing, we modify the general listing algorithm by, for each cluster $C$, not sending edges involving $C$-light nodes into $C$. Instead, we have $C$-light nodes list such $K_4$ - that is, $K_4$ that involve two nodes from $C$ and two $C$-light nodes. Notice that this is inspired by~\cite{Eden+DISC19} but is slightly different. In \cite{Eden+DISC19}, $C$-light nodes only list $K_4$ instances when \emph{both} endpoints outside the cluster are $C$-light. In our case, we have to use $C$-light nodes to list all $K_4$ instances which have an edge outside the cluster with \emph{at least one} $C$-light node incident to it. The reason for this difference is due to the fact that in \cite{Eden+DISC19}, $C$-heavy nodes send all their neighborhood into the cluster, while in our case, $C$-heavy nodes only send their \emph{outgoing} edges into the cluster, and thus it is only guaranteed that the cluster nodes know of edges between two $C$-heavy nodes, and not between a $C$-heavy and a $C$-light node. 

As we modify only the final part of the algorithm, we simply need to prove variants of Theorems~\ref{theorem:Kp_helper} and~\ref{theorem:Kp_arb} for the case $p = 4$ with a round complexity of $\tilde{O}(n^{2/3})$. The variants only omit the required condition of $\delta > 3/4$ which we no longer need, and thus the exponent of $n$ in the running time becomes $2/3$. The proof of Theorem~\ref{theorem:Kp} given the variant of Theorem~\ref{theorem:Kp_helper} remains the same, and gives a round complexity of $\tilde{O}(n^{2/3})$. Similarly, the proof of the variant of Theorem~\ref{theorem:Kp_helper} given the variant of Theorem~\ref{theorem:Kp_arb} remains the same.

The proof of the variant of Theorem~\ref{theorem:Kp_arb} is almost identical to the current proof of Theorem~\ref{theorem:Kp_arb}, with the exception of Section~\ref{subsec:bad-learn}, as the analysis of the bad edges does involve the condition $\delta>3/4$. 

Thus, we now prove that Section~\ref{subsec:bad-learn} can be replaced by an algorithm which runs in $\tilde{O}(n^{2/3})$ rounds for $K_4$, and conclude. Notice that in this proof, we do \emph{not} move edges from $E'_m$ to $\hat{E}_r$ at all.

For a cluster $C$, we set the threshold for $v\notin C$ to be a $C$-heavy node at having at least $n^{d - 1/3}$ neighbors in $C$. A node can have at most $n^d$ edge oriented away from it, and so a $C$-heavy node can in $O(n^{1/3})$ rounds send all its neighborhood into the cluster, by sending $O(n^{1/3})$ messages to each of its cluster neighbors. Notice that Remark~\ref{remark:edges} still holds since every node inside the cluster learns at most $O(n^{4/3})$ edges from outside the cluster, and since $d > 2/3$, this is at most $O(n^{d + 2/3})$ which is at most $O(n^{d + 3/4})$, as Remark~\ref{remark:edges} requires.

Notice that in an instance $K_4$ which has at least one edge in the cluster, there can be at most one edge completely outside the cluster. As such, either that edge is between two $C$-heavy nodes or it has a $C$-light node incident to it. As we sent all the edges between $C$-heavy nodes into the cluster, we can thus list inside the cluster all such instances of $K_4$ whose outside edge is between $C$-heavy nodes.

For the case of $C$-light nodes, we now perform a sequential iteration on all the clusters. Notice that there are at most $O(n^{1-\delta})$ clusters, since the clusters are node-disjoint and each has at least $n^\delta$ nodes. Iteratively, for each cluster $C$, every $v$ which is a $C$-light node, iterates on all its cluster neighbors $u$ and broadcasts $u$ to all its neighbors (in $C$ and outside $C$ alike). Once node $v$ sends to $v'$ the ID of $u$, node $v'$ responds with whether $u$ is also a neighbor of $v'$. Finally, node $v$ lists all the $K_4$ instances which it sees.

Note that the number of rounds for this procedure is $O(n^{1 - \delta + d - 1/3}) = O(n^{2/3 + d - \delta})$, which is the round complexity we aim for. 

We now show that this procedure lists all the instances of $K_4$ which have at least one edge completely in a cluster $C$ and at least one of the nodes outside of $C$ is $C$-light. In other words, we claim that given $C$, any $H = \{u, w, v, v’\}$ which is a $K_4$ instance, such that $u, w \in C$, and $v, v’ \notin C$, where $v$ is $C$-light, is listed by $v$. To prove this, recall that $v$ tells $v’$ about its neighbors $u$ and $w$, and $v’$ responds to $v$ by telling it that $u$ and $w$ are also neighbors of $v’$. Also, $v$ tells $u$ about its neighbor $w$, and $u$ responds that $w$ is also a neighbor of $u$. Thus, $v$ knows all the edges in $H$ and will list this instance of $K_4$.
\end{proof}

\vspace{-9pt}

\section{Sparsity-aware $K_p$ listing in the \clique model}
\label{section:clique}
We note that the sparsity aware algorithm executed in the \congest algorithm above can be used directly in the \clique model, as follows.

\begin{theorem-repeat}{theorem:clique}
	\cliqueTheorem
\end{theorem-repeat}

With regards to showing the upper bound, notice that the proof for the correctness and complexity of the algorithm is almost exactly the same as shown in Section~\ref{subsec:list}, with only two differences. The first is that instead of using $k^{1/p}$ parts in the partition, we use $n^{1/p}$. The second change is that in order to ensure that the requirements of Lemma~\ref{lemma:partition} are met, if $m / n^{1/p} < 20 n \log n$, then we add \emph{fake} edges to the graph until $m / n^{1/p} = 20 n \log n$. We mark these fake edges with an additional bit saying that they are fake, and thus nodes which receive them will not use them in order to list instances of $K_p$. Notice that if $m / n^{1/p} = 20 n \log n$, then our round complexity is $\tilde{O}(1)$, and so we are not hurt by adding these fake edges.

The lower bound that shows that our algorithm is tight follows directly from the lower bound proofs (for non-sparse listing) in~\cite{Fischer+SPAA18, Izumi+PODC17}, by considering a graph that contains a dense \emph{subgraph} induced by $\Theta(\sqrt{m})$ nodes.

\vspace{-3pt}

\section{Discussion}
Notice that since we solve Challenge~\ref{challenge:bringingInEdges} in $\tilde{O}(n^{3/4})$ rounds, then if it is possible to solve  Challenge~\ref{challenge:listingInsideClusters} for $K_p$ listing in $O(n^{(p-2)/p})$ rounds, then one would get an optimal algorithm for $p \geq 6$.
If, in addition, the complexity of solving Challenge~\ref{challenge:bringingInEdges} could  be brought down to $\tilde{O}(n^{1/2})$ rounds, then one would get an optimal algorithm for $p \geq 4$. Additionally, this may assist for other subgraphs apart from $K_p$. 

It is interesting that all the results in the \congest model regarding subgraph related problems with $H = K_p$ are directly for listing, and imply detection and counting algorithms with the same runtime, yet no better results are known for detection or counting for any $K_p$. In the \clique setting, $K_3$ 
is known to have a faster counting algorithm, as shown in Censor-Hillel et al.~\cite{Censor-Hillel+DC19}. 
In the \qcongest model, $K_3$ detection has a faster algorithm~\cite{Izumi+STACS20}. 

There is an inherent difficulty in attempting to apply the $K_3$ counting algorithm from the 
$\mathsf{CONGESTED}$ $\mathsf{CLIQUE}$
model to the \congest model as it involves \emph{ring matrix multiplication}, which is difficult to implement in a sparsity aware manner, even sequentially. Thus, it would be interesting if a sparsity aware algorithm for $K_p$ detection or counting in the \clique model, which would be faster than $K_p$ listing, can be developed, as such an algorithm might be implementable in the \congest model using similar techniques to what we have shown here.
~\\

\section*{Acknowledgements}
The authors would like to thank Yuval Efron and Miel Sharf for helpful discussions, Orr Fischer for elaborating upon~\cite{Eden+DISC19} and for comments about an earlier version of our paper, and Yi-Jun Chang for further elaborating upon~\cite{Chang+SODA19}.
This project has received funding from the European Union's Horizon 2020 Research And Innovation Program under grant agreement no.755839. 
FLG was supported by JSPS KAKENHI grants Nos.~JP16H01705, JP19H04066, JP20H00579, JP20H04139 and by the MEXT Quantum Leap Flagship Program (MEXT Q-LEAP) grant No.~JPMXS0118067394. 
\bibliography{References}

\end{document}